\documentclass[electronic, preprint]{vgtc}

\ifpdf
  \pdfoutput=1\relax                   
  \usepackage{graphicx}                
  \DeclareGraphicsExtensions{.pdf,.png,.jpg,.jpeg} 
\else
  \usepackage{graphicx}                
  \DeclareGraphicsExtensions{.eps}     
\fi%

\graphicspath{{figures/}{pictures/}{images/}{./}{./vgtc_conference_latex/images/}{./images/}} 

\usepackage{microtype}                 
\PassOptionsToPackage{warn}{textcomp}  
\usepackage{textcomp}                  
\usepackage{mathptmx}                  
\usepackage{times}                     
\usepackage{cite}                      
\usepackage{tabu}                      
\usepackage{booktabs}                  
\usepackage{expdlist}
\usepackage{float}

\usepackage{orcidlink}

\onlineid{0}

\vgtccategory{Research}

\vgtcinsertpkg

\preprinttext{This poster was presented at the \href{http://ieeevis.org/year/2020/welcome}{IEEE VIS 2020} poster session.}

\title{Situated Visualization in Motion}

\author{Lijie Yao \orcidlink{0000-0002-4208-5140} \thanks{e-mail: yaolijie0219@gmail.com}\\ %
        \scriptsize Université Paris-Saclay, CNRS,\\\scriptsize Inria, LRI, 91405 Orsay, France.  %
\and Anastasia Bezerianos \orcidlink{0000-0002-7142-2548} \thanks{e-mail: anastasia.bezerianos@lri.fr}\\ %
     \scriptsize Université Paris-Saclay, CNRS,\\\scriptsize Inria, LRI, 91405 Orsay, France. %
\and Petra Isenberg \orcidlink{0000-0002-2948-6417} \thanks{e-mail: petra.isenberg@inria.fr}\\ %
     \scriptsize Université Paris-Saclay, CNRS,\\\scriptsize Inria, LRI, 91405 Orsay, France. }

\teaser{
\includegraphics[width=.3\textwidth]{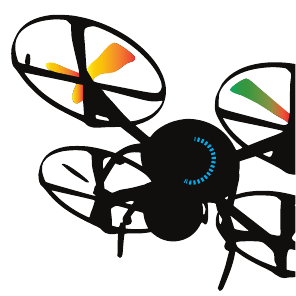}\hfill%
\includegraphics[width=.3\textwidth]{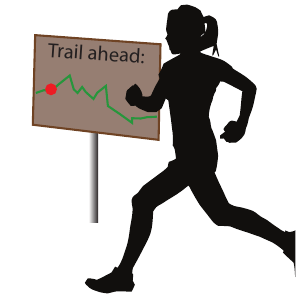}\hfill%
\includegraphics[width=.3\textwidth]{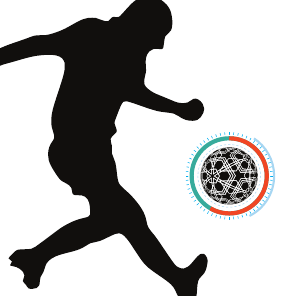}%
\caption{Three examples of visualizations viewed under motion. Left: Visualization of rotor speed and battery level attached to a flying drone. Middle: An athlete running past a sign showing the elevation levels of the trail ahead. Right: A player is hitting an augmented soccer ball.}
\label{fig:visinmotionexamples}
}

\abstract{
We contribute a first design space on \emph{visualizations in motion} and the design of a pilot study we plan to run in the fall. Visualizations can be useful in contexts where either the observation is in motion or the whole visualization is moving at various speeds. Imagine, for example, displays attached to an athlete or animal that show data about the wearer – for example, captured from a fitness tracking band; or a visualization attached to a moving object such as a vehicle or a soccer ball. The ultimate goal of our research is to inform the design of visualizations under motion. 

} 

\CCScatlist{
  \CCScatTwelve{Human-centered computing}{Visu\-al\-iza\-tion}{Visualization systems and tools}{}
}

\nocopyrightspace

\newcommand{\eg}{e.\,g.}

\begin{document}


\firstsection{Introduction}

\maketitle
Our research contributes to the area of situated data visualization \cite{Willett:2017:EDR}. In a situated data visualization, the data is directly visualized near the physical space, object, or person it refers to ($=$the data referent). Situated visualizations can often be found in contexts in which the data referent or observer does not remain static but is moving at various speeds. Imagine, for example, an athlete running with a fitness band on which he/she checks current progress or a visualization overlayed on a soccer or basketball to show game statistics. In these mobile and dynamic use cases, situated visualizations have to overcome new challenges if we want the data to be readable in real-time by the human observer. That is, we have to find effective and visually stable situated data encodings that are readable under various types, directions, and speeds of movement. In the first part of our research we are developing a design space of visualization movement types while considering the relationship to the human observer. In the continuation of our research we will explore different types of display modalities including wearable devices, augmented reality, and physical visualizations to realize concrete prototypes of situated mobile visualizations. Our next goal is to contribute empirical studies to assess how different design factors and types of movement influence the perception of situated visualizations.

\section{Related Work}
A large body of literature in the HCI community looked at how to engineer wearable devices and studied their use--sometimes under locomotion. Much of this work has focused on usability aspects and the possibility to \emph{interact}, rather than perceive information accurately. For example, Schneegass and Voit \cite{Schneegass2016} looked at the use of smart watches while in motion, analyzing how participants interacted with a GestureSleeve while running. Their usage scenario was the starting, pausing, and stopping of a fitness tracking app. 

We are not aware of studies in the Visualization community that offer any design advice for data-viewing contexts that involve motion of the whole visualization. However, there have been results for contexts in which the display itself was static but viewers were able to move. A study on the perception of visual variables on a large high-resolution wall-sized display \cite{Bezerianos:2012:POV}, for example, tested a condition in which participants were allowed to move to see information under a chosen viewing angle and distance. No improvement in accuracy of the task was found but a steep increase in task time due to movement. Another related research area is proxemic interaction in which interfaces are controlled---at least partially---by movement in front of the display (e.\,g., \cite{Badam:2016:SVE,Jakobsen:2013:IVP}). 
In summary, there is currently a lack of perceptual guidance on the efficacy of different types of visualizations and their use under motion and locomotion. This allows us to offer new research results on this emerging usage context for situated visualizations.


\section{A Design Space for Visualizations and Motion}
To describe the research space more concretely we started by deriving important dimensions to consider when we discuss visualizations that are either moving themselves or are read by a moving observer. Note, that here we do not consider visualizations where only specific marks are in motion---as is the case with most visualizations that contain animated transitions, \eg, animated scatterplots \cite{dragicevic:inria-00556177}. 

The current dimensions of our design space are:
\begin{description}[\compact]
\item[Entity in motion:] 
 This dimension includes two values that capture which entity exhibits physical motion, either the \emph{viewer} or the \emph{visualization}. For example in \autoref{fig:visinmotionexamples} the drone is in motion while a viewer controlling the drone on the ground would not be moving. The image in the middle, however contains a moving viewer and a static visualization.
\item[Characteristics of motion:]This dimension is a category of further dimensions that describe physical motion. It includes: \emph{speed, position, acceleration, direction, distance travelled, displacement, trajectory,} and \emph{time taken}. Each characteristic is divided into two subcategories: \emph{viewer} and \emph{visualization}. We discuss the characteristics of the viewer and the visualization separately, because in most cases, the state of the viewer and the visualization are not exactly the same. For example, a spectator in a soccer stadium is to a large degree static while the soccer ball carrying a visualization (\autoref{fig:visinmotionexamples}, right) constantly changes its trajectory and speed.
\item[Motion relationship between viewer \& visualization:] \hspace{-6pt}When both viewer and visualization can potentially be in motion, it is important to describe the relationship between the two types of motion. For example, a viewer sitting on an airplane watching a flight map is moving at the exact same speed as the flight map. There is very little difference in motion between both entities and we consider the relationship to be ``static.'' All examples in \autoref{fig:visinmotionexamples}, however, exhibit ``relative motion.''  
\end{description}

\section{Design Workshop}
\begin{figure}[tb]
    \centering
    \includegraphics[height=3cm]{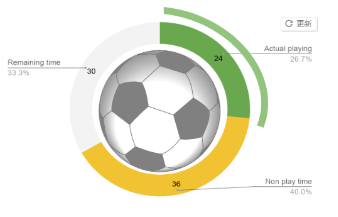}\hfill\includegraphics[height=3cm]{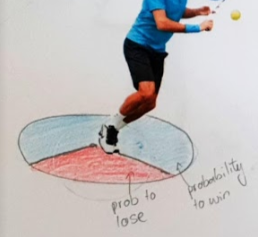}\\
    \includegraphics[height=3.05cm]{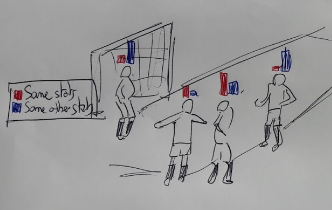}\hfill\includegraphics[height=3.05cm]{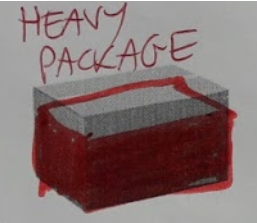}\\
    \caption{Example designs for mobile visualizations created by participants in our workshop.}
    \label{fig:designs}
\end{figure}

In order to elicit designs of visualizations in motion we conducted an online search for examples in which we found in particularly frequent examples for sports \cite{DribbleUp:2020} and visualizations on characters in online games. Next, we conducted an online design workshop with members of our visualization research team where we asked participants to come up with situated visualizations for an augmented soccer ball, augmented shipping boxes moving on a conveyor belt, and visualizations around an athlete for live sports tracking. We saw a huge variety of designs, some of which are shown in \autoref{fig:designs}. Participants used existing charts such as donut or pie charts, gauges, iconic representations such as speed lines, as well as simple text labels and graphic annotations we would not consider as data visualizations per se. Most visualizations were connected directly to the object of interest (ball, box, player) but some participants also augmented the ground / background.

Based on our analysis and assessment of the drawn and existing examples we formulated a first set of considerations for situated visualizations in motion that will serve as starting questions to answer in more concrete further empirical testing:
\begin{description}[\compact]
\item[Labels:] Moving visualizations likely require a limited set of clearly readable labels as it might be difficult to focus and refocus on different words to read while in relative motion.
\item[Unintrusive Design:] Situated visualizations in motion are add-ons to objects of primary interest: data referents such as a soccer ball, an athlete, or a game character. As such they should not take attention away from the object and blend well within the context of the referent. Visualizations directly overlayed on the object of interest might be considered too intrustive. At the same time visualizations need to be visible enough to be clearly read. 
\item[Distance to Object of Interest:] In order to establish a clear connection between data and referent, the situated visualization best stays close to the data referent and moves with it. 
\item[Simple Visualizations:] Simple known visualizations should be preferred over more complex statistical graphics that would require scanning a large area or frequent comparisons with legends.
\item[Simple Data:] Datasets of a limited number of data points and dimensions might work better for situated visualizations in motion than more complex data.
\end{description}

The list above shows that our initial investigations into the topic of situated visualizations in motion already uncovered a large number of open research questions. In particular we need to understand how complex visualizations can get when they need to be read under motion. We also need to conduct further research whether non-contiguous text such as data labels is readable read while either the viewer or the visualization is in motion. In terms of the design of the visualizations itself we also do not yet know which data visualizations can best be read under motion and if any motion characteristics may influence viewers' ability to read data correctly.

\section{Discussion and Future Work}
Currently, we are beginning to prepare an online experiment in which we want to explore how different speeds affect the readability of a simple donut chart displayed around a soccer ball. We chose the donut chart based on our initial design workshop as a frequently chosen design that has clear application possibilities in sports visualization. In the future, we want to conduct a series of further experiments based on some of the research questions outlined above. We will first focus on moving visualizations with a static viewer as this scenario was the most common in our use case scenario analysis.

\section{Acknowledgement }
This work was partly supported by the Agence Nationale de la Recherche (ANR), grant number ANR-19-CE33-0012. The authors thank for all the participants in brainstorming, they were: Mickael Sereno, Mohammad Alaul Islam, Natkamon Tovanich, Pierre Dragicevic, Tobias Isenberg, and Yuanyang Zhong.


\bibliographystyle{abbrv-doi}

\bibliography{visinmotion}

\begin{thebibliography}{1}

\bibitem{Badam:2016:SVE}
S.~K. Badam, F.~Amini, N.~Elmqvist, and P.~Irani.
\newblock Supporting visual exploration for multiple users in large display environments.
\newblock In {\em Proc.\ of the Conference on Visual Analytics Science and Technology (VAST)}, pp. 1--10. IEEE, 2016. doi: {{%
10\hspace{.1pt}\discretionary{.}{%
}{.}\hspace{.4pt}1109\discretionary{/}{%
}{/}VAST\hspace{.1pt}\discretionary{.}{%
}{.}\hspace{.4pt}2016\hspace{.1pt}\discretionary{.}{%
}{.}\hspace{.4pt}7883506}}


\bibitem{Bezerianos:2012:POV}
A.~Bezerianos and P.~Isenberg.
\newblock Perception of visual variables on tiled wall-sized displays for information visualization applications.
\newblock {\em IEEE Transactions on Visualization and Computer Graphics}, 18(12):2516--2525, Dec. 2012. doi: {{%
10\hspace{.1pt}\discretionary{.}{%
}{.}\hspace{.4pt}1109\discretionary{/}{%
}{/}TVCG\hspace{.1pt}\discretionary{.}{%
}{.}\hspace{.4pt}2012\hspace{.1pt}\discretionary{.}{%
}{.}\hspace{.4pt}251}}


\bibitem{dragicevic:inria-00556177}
P.~Dragicevic, A.~Bezerianos, W.~Javed, N.~Elmqvist, and J.-D. Fekete.
\newblock Temporal distortion for animated transitions.
\newblock In {\em Proc.\ of the Conference on Human Factors in Computing Systems (CHI)}, pp. 2009--2018. ACM, May 2011. doi: {{%
10\hspace{.1pt}\discretionary{.}{%
}{.}\hspace{.4pt}1145\discretionary{/}{%
}{/}1978942\hspace{.1pt}\discretionary{.}{%
}{.}\hspace{.4pt}1979233}}


\bibitem{DribbleUp:2020}
{DribbleUp, Inc.}
\newblock Welcome to {DU}.
\newblock Website.
\newblock https://dribbleup.com/.

\bibitem{Jakobsen:2013:IVP}
M.~Jakobsen, Y.~S. Haile, S.~Knudsen, and K.~Hornb{\ae}k.
\newblock Information visualization and proxemics: Design opportunities and empirical findings.
\newblock {\em IEEE Transactions on Visualization and Computer Graphics}, 19(12):2386--2395, 2013.

\bibitem{Schneegass2016}
S.~Schneegass and A.~Voit.
\newblock Gesturesleeve: Using touch sensitive fabrics for gestural input on the forearm for controlling smartwatches.
\newblock In {\em Proceedings of the International Symposium on Wearable Computers}, pp. 108--115. ACM, 2016. doi: {{%
10\hspace{.1pt}\discretionary{.}{%
}{.}\hspace{.4pt}1145\discretionary{/}{%
}{/}2971763\hspace{.1pt}\discretionary{.}{%
}{.}\hspace{.4pt}2971797}}


\bibitem{Willett:2017:EDR}
W.~Willett, Y.~Jansen, and P.~Dragicevic.
\newblock Embedded data representations.
\newblock {\em IEEE Transactions on Visualization and Computer Graphics}, 23(1):461--470, Jan. 2017. doi: {{%
10\hspace{.1pt}\discretionary{.}{%
}{.}\hspace{.4pt}1109\discretionary{/}{%
}{/}TVCG\hspace{.1pt}\discretionary{.}{%
}{.}\hspace{.4pt}2016\hspace{.1pt}\discretionary{.}{%
}{.}\hspace{.4pt}2598608}}


\end{thebibliography}
\end{document}